\documentclass[lettersize,journal]{IEEEtran}
\usepackage{amsmath,amsfonts}
\usepackage{algorithmic}
\usepackage{algorithm}
\usepackage{array}
\usepackage[caption=false,font=normalsize,labelfont=sf,textfont=sf]{subfig}
\usepackage{textcomp}
\usepackage{stfloats}
\usepackage{url}
\usepackage{verbatim}
\usepackage{graphicx}
\usepackage{cite}
\usepackage{tabularx,booktabs} 
\usepackage{multirow}
\usepackage{bbm}
\usepackage[dvipsnames]{xcolor}
\usepackage{hyperref}  
\begin{document}
\title{\LARGE Energy Efficiency Optimization for Subterranean LoRaWAN Using A Reinforcement Learning Approach: A Direct-to-Satellite Scenario}

\author{Kaiqiang Lin,~\IEEEmembership{Student Member,~IEEE,}
        Muhammad~Asad~Ullah,~\IEEEmembership{Student~Member,~IEEE,}
        Hirley~Alves,~\IEEEmembership{Member,~IEEE,}
        Konstantin Mikhaylov,~\IEEEmembership{Senior Member,~IEEE,}
        and Tong~Hao,~\IEEEmembership{Member,~IEEE}
\thanks{K. Lin and T. Hao are with the College of Surveying and Geo-Informatics, Tongji University, Shanghai, China. E-mail: lkq1220@tongji.edu.cn; tonghao@tongji.edu.cn. (\textit{Corresponding Author: Tong Hao})}
\thanks{M. Asad Ullah, H. Alves and K. Mikhaylov are with the Centre for Wireless Communications, University of Oulu, Finland. M. Asad Ullah is also with VTT Technical Research Centre of Finland Ltd, Oulu, Finland. Email: Muhammad.AsadUllah@oulu.fi; Hirley.Alves@oulu.fi;  konstantin.mikhaylov@oulu.fi.}
\thanks{This work was supported in part by National Natural Science Foundation of China (No. 42211530077 and 42074179), the Academy of Finland, 6G Flagship program (No. 346208), and the China Scholarship Council.}~\vspace{-10mm}
}
\maketitle

\begin{abstract}
The integration of subterranean LoRaWAN and non-terrestrial networks (NTN) delivers substantial economic and societal benefits in remote agriculture and disaster rescue operations. The LoRa modulation leverages quasi-orthogonal spreading factors (SFs) to optimize data rates, airtime, coverage and energy consumption. However, it is still challenging to effectively assign SFs to end devices for minimizing co-SF interference in massive subterranean LoRaWAN NTN. To address this, we investigate a reinforcement learning (RL)-based SFs allocation scheme to optimize the system's energy efficiency (EE). To efficiently capture the device-to-environment interactions in dense networks, we proposed an SFs allocation technique using the multi-agent dueling double deep Q-network (MAD3QN) and the multi-agent advantage actor-critic (MAA2C) algorithms based on an analytical reward mechanism. Our proposed RL-based SFs allocation approach evinces better performance compared to four benchmarks in the extreme underground direct-to-satellite scenario. Remarkably, MAD3QN shows promising potentials in surpassing MAA2C in terms of convergence rate and EE.  
\end{abstract}

\vspace{-1mm}
\vspace{-6mm}
\section{Introduction}
\IEEEPARstart{T}{he} integration of LoRaWAN-based wireless underground sensor networks and non-terrestrial networks (NTN) enables subterranean massive machine-type communications (mMTC) applications to operate in hard-to-reach or disaster rescue areas~\cite{LinUtSMag}. LoRa, a chirp spread spectrum modulation variation in LoRaWAN, introduces quasi-orthogonality between packets with different spreading factors (SFs)~\cite{LoRaWCL}. This characteristic grants LoRa its resistance to interference while offering a range of trade-offs between time-on-air (ToA), radio coverage, and energy consumption through varying SF levels. However, in subterranean mMTC scenarios, the Aloha-like media access protocol used in LoRa constrains the network capacity and collision robustness. For instance, the simulation results reported in~\cite{LinUtSMag, AsadTII} illustrate a relatively low probability of successful packet delivery when a large number of end devices (EDs) are assigned with the same SF in the underground direct-to-satellite (U-DtS) connectivity. This is attributed to the frequent co-SF interference that occurs when packets featuring the same SF are simultaneously transmitted on the same channel. 

To leverage LoRa quasi-orthogonality, several studies have discussed the SFs allocation techniques and evaluated the scalability for terrestrial networks. Specifically, two notable one-time spatial SFs allocation schemes, namely equal-interval-based (EIB)~\cite{CaptureeffectTII} and equal-area-based (EAB)~\cite{EAB}, were proposed to mitigate the co-SF interference and improve the packet delivery ratio. To adjust to dynamic underground environments, adaptive parameters assignment schemes were proposed in LoRaWAN, such as the adaptive data rate mechanism specified by the LoRa Alliance~\cite{EPPIoT} and the path-loss-based (PLB) scheme proposed in~\cite{PLB}. Nevertheless, both solutions overlook the co-SF interference. Thus, their performance diminishes in practical subterranean mMTC applications. Recently, reinforcement learning (RL) has shown to be a promising paradigm for solving the SFs allocation problem in LoRaWAN. For instance, in~\cite{LoRaDRL1, LoRaDRL2}, authors used a single-agent RL (SARL) approach to derive the optimal SFs allocation by considering the co-SF interference for improving the network reliability and throughput. To further enhance the exploration efficiency of SARL in mMTC applications, a multi-agent RL (MARL) approach has been applied in~\cite{MADLWUSN} to determine the optimal SFs allocation for improving the energy efficiency of underground EDs. However, the above RL approaches only adopt the basic deep Q-network (DQN). Considering the issues related to the overestimation and imprecision of Q value in the basic DQN, the multi-agent dueling double DQN (MAD3QN) is proposed to augment the agents' optimization capabilities~\cite{D3QN}. Meanwhile, in~\cite{A2C}, another mainstream MARL algorithm based on value-based and policy-based optimization, namely the multi-agent advantage actor-critic (MAA2C), is developed to provide a more efficient exploration strategy compared to DQN.

To the best of our knowledge, there have been no studies exploring the effectiveness of MAD3QN or MAA2C in optimizing SFs allocation in LoRaWAN, let alone our considered massive subterranean NTN scenarios. Notably, energy efficiency is a significant metric from both economic and sustainable perspectives for the design of such a system~\cite{R1ref1,R1ref2}. Motivated by this, this letter utilizes the MAD3QN and the MAA2C algorithms with our developed analytical reward mechanism for SFs allocation, aiming to maximize the system's energy efficiency characterized by the average amount of energy consumed for uplink packet delivery. The simulation results demonstrate the superiority of our approach over four well-established benchmarks in extreme U-DtS scenarios.

\vspace{-4mm}
\section{System Model}
\begin{figure}[t]
    \centering
    \includegraphics[width=2.2in]{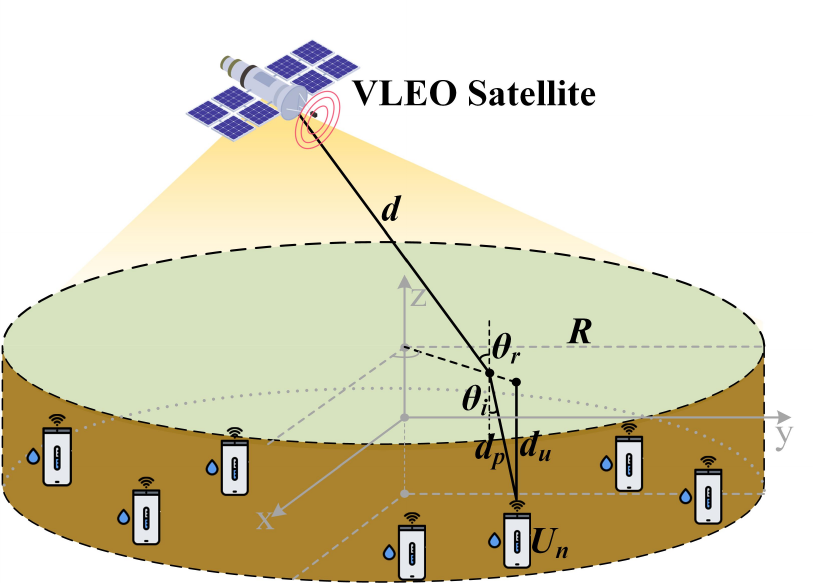}
    \vspace{-3mm}
    \caption{The subterranean LoRaWAN NTN system taking a VLEO satellite as an example.}
    \label{fig_sys}
    \vspace{-6mm}
\end{figure}
For the sake of clarity, in the rest of the paper, we focus on a U-DtS example based on a very-low-Earth-orbit (VLEO) satellite. However, the presented methods and obtained results can be generalized for any underground-to-NTN connectivity scenario, including those employing unmanned aerial vehicles or high-attitude platforms. Consider the subterranean LoRaWAN NTN system depicted in Fig.~\ref{fig_sys} where a LoRaWAN gateway (GW) deployed on the non-terrestrial (NT) platform generates a single spot beam for covering a massive set $\mathcal{U}=\{U_{n}|n=1,2,\dots, N\}$ of underground EDs buried at the same depth $d_u$. Specifically, $N$ underground EDs are distributed according to the Poisson point process (PPP) with intensity $\lambda=\frac{N}{\pi R^2 T_p}$ within a circular region of radius $R$. Herein, we assume that all EDs transmit an uplink packet with a period of $T_p$, the fixed physical layer (PHY) payload $PL$, the same bandwidth $B$, and the maximum transmit power $P_t$. A LoRa packet is successfully demodulated when the received signal-to-noise ratio (SNR) and signal-to-interference ratio (SIR) requirements are satisfied.

\subsubsection{Path Loss Model} The total path loss from $U_n$ to the NT GW comprises the air attenuation $L_{air}$, the fraction loss on the soil-air interface $L_r$, and the attenuation in underground soil $L_{soil}$. Consider that a ground-to-satellite link propagates in free space and the attenuation caused by the ionosphere, atmospheric gases, fog, clouds, and rain droplets can be neglected in the sub-GHz frequency band, the free space path loss model is adopted for the air path loss calculation~\cite{LinUtSMag}. Hence, the total path loss model is given by~\cite{LinLoRaWUSNs, LinUtSMag}.
\vspace{-2mm}
\begin{equation}
\label{EqPL}
    g(d) =\! L_{air}L_{r}L_{soil} \!= \left(\frac{4 \pi f_c}{c}\right)^2 \left(d \right)^\eta \left(\frac{2 \beta d_{p}}{\exp(-\alpha d_{p})}\right)^{2}, 
    \vspace{-2mm}
\end{equation}
where $f_c$ is the carrier frequency, $c$ denotes the speed of light in free space, $d$ is the distance between $U_n$ and the NT GW, $\eta$ is the path-loss exponent, $d_{p}=d_u/(\cos{\arcsin({1/\sqrt{\varepsilon'}}}))$ is the length of the underground path; $\alpha$ and $\beta$ are the attenuation constant and phase shifting constant, respectively, 
\vspace{-2mm}
\begin{align}
    \alpha &= 2 \pi f_c \sqrt{\frac{\mu_{r} \mu_{0} \varepsilon' \varepsilon_{0}} {2}[\sqrt{1+(\varepsilon''/{\varepsilon'})^{2}}-1 ]}, \\
    \beta &= 2 \pi f_c \sqrt{\frac{\mu_{r} \mu_{0} \varepsilon' \varepsilon_{0}} {2}[\sqrt{1+(\varepsilon''/{\varepsilon'})^{2}}+1 ]},
    \vspace{-2mm}
\end{align}
where $\mu_{r}$ is the soil’s relative permeability, $\mu_{0}$ is the free-space permeability, $\varepsilon_{0}$ is the free space permittivity. At the same time, $\varepsilon'$ and $\varepsilon''$ are the real and imaginary parts of the soil’s complex dielectric constant (CDC), i.e., $\varepsilon = \varepsilon' + j\varepsilon''$. CDC can be calculated by the mineralogy-based soil dielectric model developed in~\cite{MBSDM}. Notably, the refraction loss on the soil-air interface $L_{r}$ can be neglected in our study, implying $L_{r}=1$. This is because most energy is refracted when electromagnetic waves propagate from a high-density medium (soil) to a lower-density one (air).
\subsubsection{Success Probability for SNR Guarantee} In the absence of interference, the probability of successfully decoding a packet as a function of distance $d$ is
\vspace{-2mm}
\begin{equation}
\label{eqPSNR}
    P_{SNR}(d) \!= \! \mathbb{P}\left[\frac{P_t G_t G_r |h|^2}{g(d) \sigma_w^2} > q \right] \!=\! \exp\left(-\frac{g(d) q \sigma_w^2}{P_t G_t G_r}\right),
    \vspace{-2mm}
\end{equation}
where $G_t$ and $G_r$ are the antenna gains of the underground ED and the NT GW, respectively, $|h|^2$ accounts for fading in the EDs-to-GW channel, whose coefficients are characterized by Rayleigh fading and the power follows an exponential distribution with a unit mean, $\sigma_w^2$ is the variance of the additive white Gaussian noise, and LoRa SF-specific SNR demodulation threshold $q=\{-6, -9, -12, -15, -17.5, -20\}$~dB for SF7$\sim$12 denoted by $\{SF_k| k =1, \ldots, 6\}$, respectively~\cite{LoRaWCL}.

\subsubsection{Success Probability for SIR Guarantee} The recent studies have demonstrated the presence of the capture effect for LoRa signals, which implies that a receiver demodulates the stronger packet under the interference of the weaker ones if the SIR is above a certain threshold~\cite{CaptureeffectTII}. Given the SIR threshold $\delta$ and the interference set (i.e., simultaneously transmitted packets featuring the same-SF) $\Phi$, the SIR success probability according to distance $d$ is
\vspace{-2mm}
\begin{align}
\label{eqPSIR}
    P_{SIR}(d) &= \mathbb{P} \left[\frac{|h|^2 g(d)}{\sum_{i \in \Phi}\left|h_i\right|^2 g(d_i)} > \delta \right] \nonumber \\
    & \stackrel{(a)}{=} \exp\!\left(\frac{-4 N_k \cdot ToA_{k}}{d_{max}^2 T_p N_c} \int_0^{d_{max}} \!\!\! \frac{\delta d^\eta d_i^{-\eta}}{1+\delta d^\eta d_i^{-\eta}} d_i \mathrm{~d} d_i\right) \nonumber \\
    & \stackrel{(b)}{=}\exp\! \left[\frac{-2 N_k \cdot ToA_{k}}{T_p N_c} { }_2 F_1\!\left(1, \frac{2}{\eta} ; 1+\frac{2}{\eta} ;-\frac{d_{max}^{\eta}}{\delta d^\eta}\right)\!\right], 
    \vspace{-2mm}
\end{align}
where (a) follows after using the probability generating functional of the product over PPP~\cite{CaptureeffectTII}, and (b) is obtained by adopting the definition of the Gauss Hypergeometric function ${ }_2 F_1(\cdot)$~\cite{JeanWCL}. Furthermore, $i$ represents the interfering signal, $N_k$ is the number of EDs assigned by $SF_k$, $d_{max}$ denotes the maximum distance between EDs and the NT GW, $ToA_{k}$ is the time-on-air of $SF_k$, and $N_c$ is the number of uplink channels.

\subsubsection{Packet Delivery Ratio} 
\begin{figure}[t]
    \centering
    \includegraphics[width=2.4in]{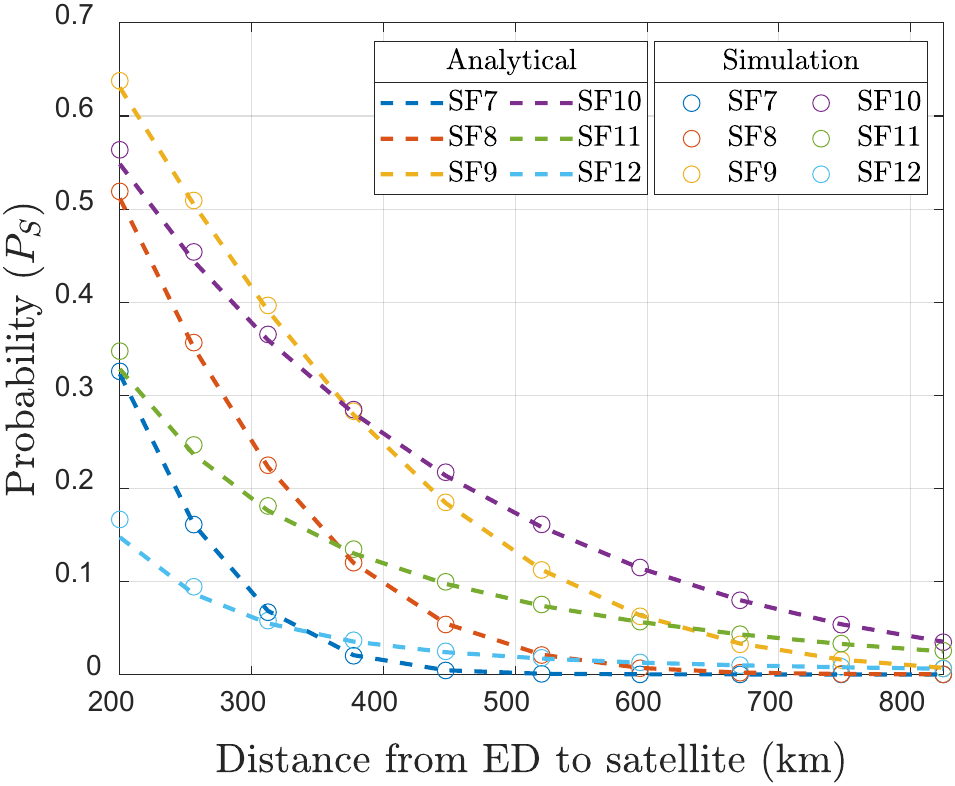}
    \vspace{-3mm}
    \caption{Analytical and simulated overall success probability $P_S$ versus distance from ED to satellite $d$. The related parameters can be found in~Table~\ref{tab}.}
    \label{fig_ps}
    \vspace{-6mm}
\end{figure}

The overall probability of successful packet delivery is the product of $P_{SNR}$ and $P_{SIR}$
\vspace{-2mm}
\begin{equation}
\label{eqPS}
    {P_S}(d) = P_{SNR}(d)P_{SIR}(d).
    \vspace{-2mm}
\end{equation}
Fig.~\ref{fig_ps} highlights that the success probability $P_S$ of the analytical model described in~\eqref{eqPS} agrees well with that obtained from the Monte-Carlo simulations, where $N=1000$ (i.e., 1k) EDs transmit a 23-byte PHY payload packet in a single uplink channel with the period of $T_p=600$~s. 

\subsubsection{Energy Per Packet (EPP)} The system's energy efficiency is characterized by EPP, which denotes the average amount of energy consumed by an ED to successfully deliver a packet to the NT GW~\cite{EPPIoT}. Note that for the sake of tractability, we do not consider the downlink communication (i.e., LoRaWAN receive windows) and the energy consumption associated with it. Thus, EPP is given by
\vspace{-2mm}
\begin{equation}
\label{eqEPP}
    EPP = \frac{V_{supply} I_{tx} ToA_{k}} {P_S},
    \vspace{-2mm}
\end{equation}
where $V_{supply}=3.3$~V is the supply voltage of the ED, while $I_{tx}$ is the transmit current consumption determined by $P_t$. 

\vspace{-2mm}
\section{RL-Based SFs allocation Approach}
This work aims to determine the optimal SFs allocation strategy that minimizes the system's EPP in massive U-DtS scenarios. Hence, the optimization objective is formulated as 
\vspace{-2mm}
\begin{equation}
\label{P1}
{\mathop {\min} \limits_{\{ {a_1, \ldots, a_N}\} }} \frac{1}{N}\sum_{n=1}^N EPP_n~~s.t.~a_n \in \{SF_k\},
\vspace{-2mm}
\end{equation}
where $a_n$ is the selected SF configuration for the $n$-th ED. To achieve this, we utilize two MARL approaches, i.e., MAD3QN and MAA2C. Concretely, we consider each ED as an independent RL agent responsible for selecting and utilizing an SF configuration for packet transmission. The NT GW then applies the MARL approach to derive the SFs allocation strategy based on the received information from all agents. Finally, the NT GW broadcasts the SFs allocation results to all EDs. The MARL components are as follows: 
\subsubsection{Agent} Each agent includes $(s_n^t; a_n^t; r_n^t; s_n^{t+1})$, which implies the $n$-th ED in state $s_n^t$ chooses an action $a_n^t$ according to certain policy at step $t$. Then, it will receive its own $r_n^t$ and fall into state $s_n^{t+1}$ at the next training step.

\subsubsection{Action Space $\mathcal{A}$} 
The action space of each agent is $A = \{SF_k\}$. Thus, the selected actions of all agents at step $t$ can be defined as $\mathcal{A}^t = \{a_1^t, \ldots, a_N^t\}, a \in A$.  

\subsubsection{State Space $\mathcal{S}$} 
The state observed in each agent consists of the selected SF configuration, SNR probability, SIR probability, packet delivery ratio, and EPP, which can be denoted as $s_n^t = \{a_n^t, (P_{SNR})_n^t, (P_{SIR})_n^t, (P_{S})_n^t, EPP_n^t\}$. Hence, the state set of all agents at step $t$ is $\mathcal{S}^t = \{s_1^t, \ldots, s_N^t\}$. 

\subsubsection{Reward $\mathcal{R}$}
The objective of the RL approach is to minimize the EPP; thus, the reward of each agent at step $t$ is defined as $r_n^t = \frac{1}{EPP_n^t}$. Consequently, the reward set of all agents is $\mathcal{R}^t = \{r_1^t, \ldots, r_N^t\}$. Notably, our reward mechanism, which accounts for local and global rewards, contributes to the expeditious convergence of the MARL algorithm.

\vspace{-4mm}
\subsection{MAD3QN Approach}
The workflow of the MAD3QN approach is illustrated in \textbf{Algorithm~\ref{algd3qn}}. The value-based MAD3QN is devoted to obtaining an optimal policy, which maps a state to a distribution over actions~\cite{D3QN}. Each agent's policy is characterized by the DQN $Q(s_n^t, a; \omega_n)$ to estimate the expectation of action-value distribution, where $\omega_n$ denotes the policy network weights of the $n$-th agent. Hence, MAD3QN aims to search for optimal weights of each agent by minimizing the loss function, i.e.,
\vspace{-2mm}
\begin{align}
\label{EqD3QNLoss}
y_n^t &= r_n^t\!+\!\gamma Q(s_n^{t+1}, \underset{a\in A}{\mathrm{argmax}}~Q(s_n^{t+1}, a; \omega_n); \hat{\omega}_n), \nonumber \\
\hat{L_n} &= (y_n^t - Q(s_n^t, a_n^t; \omega_n))^2,
\vspace{-2mm}
\end{align}
where $\gamma \in [0, 1)$ denotes a discount factor that balances the trade-off between immediate and future rewards, and $\hat{\omega}_n$ denotes the target network weights of the $n$-th agent. The target network is generated by cloning the current network and updating the weights after a fixed number of iterations. The network weight $\omega_n$ is updated through a gradient descent method, i.e., $\omega_n = \omega_n - \tau \nabla_{\omega_n} \hat{L_n}(\omega_n)$, where $\tau$ is the learning rate. Furthermore, compared with the single estimator in the basic DQN, the output of each agent in MAD3QN is divided into two estimators, i.e., value function and advantage function, to accelerate the convergence. Accordingly, the Q-value of each agent in MAD3QN can be presented as
\vspace{-2mm}
\begin{align}
\label{EqduelingQ}
& Q\left(s_n^t, a_n^t ; \bar{\omega}_n, \kappa_n, \nu_n \right) = \mathbb{V}\left(s_n^t; \bar{\omega}_n, \kappa_n \right)+ \nonumber \\
& \quad\left[\mathbb{A}\left(s_n^t, a_n^t; \bar{\omega}_n, \nu_n \right)-\frac{1}{|A|} \sum_{a \in A} \mathbb{A}\left(s_n^t, a; \bar{\omega}_n, \nu_n \right)\right],
\vspace{-2mm}
\end{align}
where $\bar{\omega}_n, \kappa_n, \nu_n$ are the weights of the shared convolutional encoder, value function $\mathbb{V}(\cdot)$ and advantage function $\mathbb{A}(\cdot)$, respectively, for the $n$-th agent. 

\begin{algorithm}[t]
\small
\caption{MAD3QN Approach}
\label{algd3qn}
\renewcommand{\algorithmicrequire}{\textbf{Input:}}  
\renewcommand{\algorithmicensure}{\textbf{Output:}} 
\begin{algorithmic}[1] 
\STATE Initialize initial state $\mathcal{S}^0$, policy network $Q(s_n^0, a; \omega_n)$ with random weights $\omega_n$, target Q-network $Q(s_n^0, a; \hat{\omega}_n)$ with $\hat{\omega}_n =\omega_n$, replay memory $\mathcal{M}$, $\epsilon_{t}=0$, $\epsilon_{T}=0.9999$, $\gamma=0.7$, and $m=100$    
    
\FOR{$t = 1$ \TO $T_{max}$}
    \FOR{$n=1$ to $N$}
        \STATE $a_n^t = \left\{\begin{aligned} & \text{Randomly select}~a_n^t \in A, \mathrm{rand()}>\epsilon_t \\ &\underset{a\in A}{\mathrm{arg max}}~Q(s_n^t, a; \omega_n), \text{otherwise} \end{aligned} \right.$
    \ENDFOR
    \STATE Execute action $\mathcal{A}^t$ in the environment and get $\left(\mathcal{S}^t,\mathcal{A}^t,\mathcal{S}^{t+1}, \mathcal{R}^t\right)$ by~\eqref{eqPSNR}, \eqref{eqPSIR}, \eqref{eqPS}, and \eqref{eqEPP}
    \FOR{$n=1$ to $N$}
        \STATE Store transition $\left(a_n^t,s_n^t,r_n^t, s_n^{t+1}\right)$ in $\mathcal{M}$
        \IF{$\mathcal{M}$ is full}
            \STATE Sample random mini-batch of transitions from $\mathcal{M}$
            
            \STATE  Update $\omega_n$ by performing a gradient descent step on \eqref{EqD3QNLoss} 
        \ENDIF  
        \STATE Set state $s_n^t = s_n^{t+1}$
        \STATE Update $\epsilon_t$ with $\epsilon_t=\min(\epsilon_{T}, \epsilon_t+0.0002)$
        \STATE Every $m$ steps clone $\omega_n$ to $\hat{\omega}_n$
    \ENDFOR
\ENDFOR
\ENSURE Learned $Q(s_n^t, a; \omega_n)$

\end{algorithmic}
\end{algorithm}
\vspace{-4mm}

\subsection{MAA2C Approach}
The workflow of the MAA2C approach is described in \textbf{Algorithm~\ref{alga2c}}. Unlike MAD3QN, MAA2C focuses on training the critic function $C(s_n^{t}; \hat{\psi_n})$ that measures average expected return from current state $s_n^{t}$ to obtain the optimal actor policy $T(a_n^t|s_n^t; {\psi_n})$ of the $n$-th agent~\cite{A2C}. The $\psi_n$ and $\hat{\psi_n}$ are the actor and critic network weights, respectively. MAA2C aims to obtain the optimal policy of each agent by minimizing the loss of actor and critic functions. The loss function of the critic network for the $n$-th agent is  
\vspace{-2mm}
\begin{align}
\label{Eqcritic}
    z_n &= r_n^t + \gamma C(s_n^{t+1}; \hat{\psi_n}) - C(s_n^{t}; \hat{\psi_n}), \nonumber \\
    \hat{L_C} &= \left(z_n\right)^2.
    \vspace{-2mm}
\end{align}
Meanwhile, the loss function of the actor network is 
\vspace{-2mm}
\begin{equation}
\label{Eqactor}
    \hat{L_T} = -z_n\log(T(a_n^t|s_n^t; {\psi_n})). 
    \vspace{-2mm}
\end{equation}
Herein, the network weights of actor and critic are updated by a gradient descent method, i.e., $\psi_n = \psi_n-\tau \nabla_{\psi_n} \hat{L_T}(\psi_n)$ and $\hat{\psi_n} = {\hat{\psi_n}}-\tau \nabla_{\hat{\psi_n}} \hat{L_C}(\hat{\psi_n})$, respectively. 

\begin{algorithm}[t]
\small
\caption{MAA2C Approach}
\label{alga2c}
\renewcommand{\algorithmicrequire}{\textbf{Input:}}  
\renewcommand{\algorithmicensure}{\textbf{Output:}} 
\begin{algorithmic}[1]
\STATE Initialize  initial state $\mathcal{S}^0$, actor networks $T(a_n^0|s_n^0; \psi_n)$ with random weights $\psi_n$, critic network $C(s_n^0; \hat{\psi}_n)$ with random weights $\hat{\psi}_n$, and $\gamma=0.7$ 
\FOR{$t = 1$ \TO $T_{max}$}
    \FOR{$n=1$ to $N$}
        \STATE Select  $a_n^t \sim T(a_n^t|s_n^t; \psi_n)$
    \ENDFOR
        \STATE Execute action $\mathcal{A}^t$ in the environment and get $\left(\mathcal{S}^t,\mathcal{A}^t,\mathcal{S}^{t+1}, \mathcal{R}^t\right)$ by~\eqref{eqPSNR}, \eqref{eqPSIR}, \eqref{eqPS}, and \eqref{eqEPP}
    \FOR{$n=1$ to $N$}
        \STATE Update $\hat{\psi}_n$ by performing a gradient descent step on~\eqref{Eqcritic} 
        \STATE Update $\psi_n$ by performing a gradient descent step on~\eqref{Eqactor} 
        \STATE Set state $s_n^t = s_n^{t+1}$
    \ENDFOR
\ENDFOR
\ENSURE Learned $T(a_n^t|s_n^t; \psi_n), C(s_n^t; \hat{\psi}_n)$
\end{algorithmic}
\end{algorithm}
\vspace{-3mm}

\section{Simulation Results and Analysis}

\begin{table}[t]
\caption{Simulation Parameters} 
\label{tab}
\centering
\vspace{-2mm}
\begin{tabular}{m{0.27\textwidth}<{\raggedright} m{0.15\textwidth}<{\centering}}%
\toprule
\textbf{Parameters}                & \textbf{Values}                    \\  \hline
\multicolumn{2}{l}{\textbf{Operation Environments}~\cite{LinUtSMag}}\\ \hline
Total nodes ($N$)                  &1k$\sim$10k                          \\
Burial depth ($d_u$)                 & 0.4~m                               \\
VWC ($m_v$)   & 11.19\(\%\)                         \\
Clay ($m_c$)                         & 16.86\(\%\)                         \\
PHY payload ($PL$)                 & 23~Bytes                            \\
Report period ($T_p$)                & 600~s                               \\ 
Traffic pattern                    & Periodic                            \\ \hline
\multicolumn{2}{l}{\textbf{VLEO Satellite Configuration}~\cite{VLEOref}}       \\ \hline
Elevation angles ($E$)             & \(10^\circ \leq E \leq 90^\circ\)   \\
Orbital height ($H$)               & 200~km                              \\
Link distance ($d$)                & 200~km$\sim$825~km                  \\ 
Coverage radius ($R$)              & 822~km                               \\\hline
\multicolumn{2}{l}{\textbf{Radio Configuration}~\cite{LinUtSMag}}                         \\ \hline
Carrier frequency ($f_{c}$)        & 915~MHz                             \\ 
Antenna gains ($G_t$ and $G_r$)    & (2.15~dBi, 35~dBi)                 \\
SIR threshold (\(\delta\))         &6~dB                                 \\  
Path  loss exponent ($\eta$)       & 2                                  \\
Uplink channel number ($N_c$)      & 1                                  \\
Transmit power ($P_{t}$)          & 20~dBm                              \\
Transmit current ($I_{tx}$)        & 133~mA                              \\
BW ($B$)                           & 125~KHz                             \\ 
\end{tabular}
\begin{tabular}{m{0.11\textwidth}<{\raggedright} m{0.08\textwidth}<{\centering} m{0.1\textwidth}<{\centering} m{0.08\textwidth}<{\centering}}
\hline
\multicolumn{4}{l}{\textbf{Agent Setting}}                                                      \\ \hline
\multirow{2}{*}{Approach} & \multirow{2}{*}{MAD3QN} & \multicolumn{2}{c}{MAA2C}  \\ \cline{3-4} 
                          &                         & Actor        & Critic      \\ \hline
Learning rate ($\tau$)    & 0.001                   & 0.001        & 0.001       \\ \cline{2-4} 
Input layer              & linear, $|s_n|$         & linear, $|s_n|$  & linear, $|s_n|$ \\
Hidden layer             & ReLU, 16                & ReLU, 16         & ReLU, 16    \\
Output layer             & linear, $|A|$         & softmax, $|A|$ & linear, 1   \\ 
\bottomrule
\end{tabular}
\vspace{-5mm}
\end{table}

\subsection{Simulation Settings}
To evidence the performance of our proposed approach, the extreme subterranean LoRaWAN NTN scenario, i.e., U-DtS, is considered in the simulation. Herein, we envision a LoRaWAN GW deployed on a VLEO satellite, which provides one spot beam covering the real-life center-pivot irrigation farm~\cite{LinUtSMag, VLEOref}. The specific simulation parameters are listed in Table~\ref{tab}. To ensure the execution of the algorithm, we assume that our system is capable of compensating for Doppler effects and accurately predicting the positions of satellites~\cite{Assumption}. For the MARL parameters, we set that the reply memory size is $|\mathcal{M}|=16$, the mini-batch size is 4, and the maximum training episode is $T_{max}=6000$. We optimize the weights of all DQNs with the Adam optimizer. Besides, the hyper-parameter settings, the activation functions, and the neuron numbers of each deep neural network layer are summarized in~Table~\ref{tab}.

For performance comparison, four benchmarks are implemented: (1) Same SF~\cite{AsadTII}: all EDs are assigned by the same SF; (2) EIB~\cite{CaptureeffectTII}: the coverage area is partitioned into six equal-width (i.e., $\frac{R}{|A|}$) annuli for SFs allocation; (3) EAB~\cite{EAB}: the coverage area is segmented into six equal-area (i.e., $\frac{\pi R^2}{|A|}$) annuli for SFs allocation; (4) PLB~\cite{PLB}: annuli are determined based on the path loss model described in~\eqref{EqPL} and the SF-specific SNR threshold $q$.

\vspace{-4mm}

\subsection{Simulation Results}
\begin{figure}[t]
    \centering
    \includegraphics[width=2.4in]{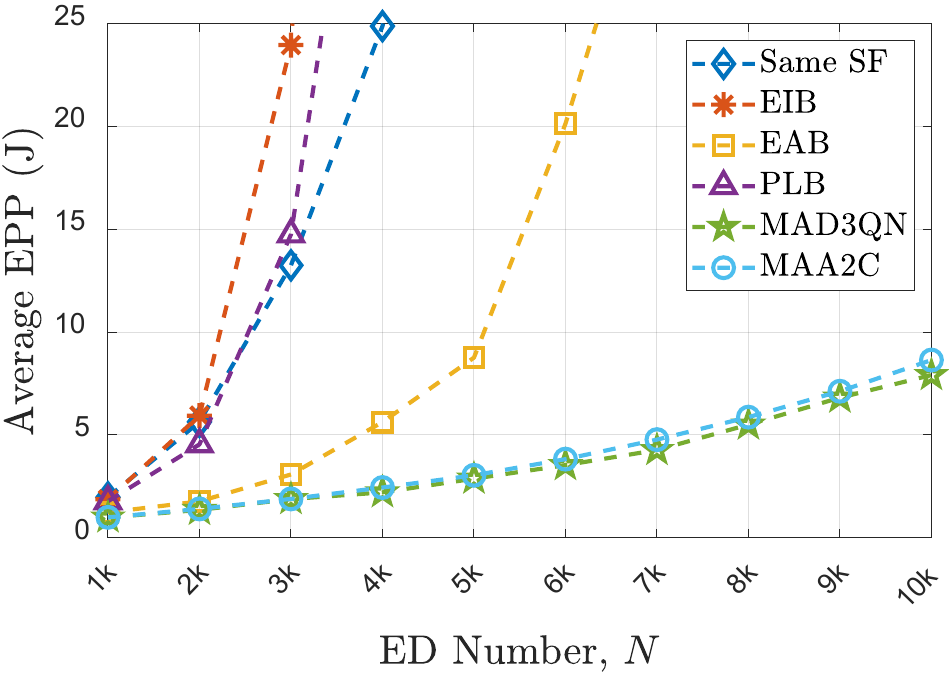}
    \vspace{-3mm}
    \caption{The comparison of the average EPP versus the number of EDs for proposed RL-based SFs allocation scheme and the same SF~\cite{AsadTII}, EIB~\cite{CaptureeffectTII}, EAB~\cite{EAB}, and PLB~\cite{PLB} based techniques.}
    \label{fig_epp}
    \vspace{-4mm}
\end{figure}

\begin{figure}[t]
    \centering
    \includegraphics[width=2.4in]{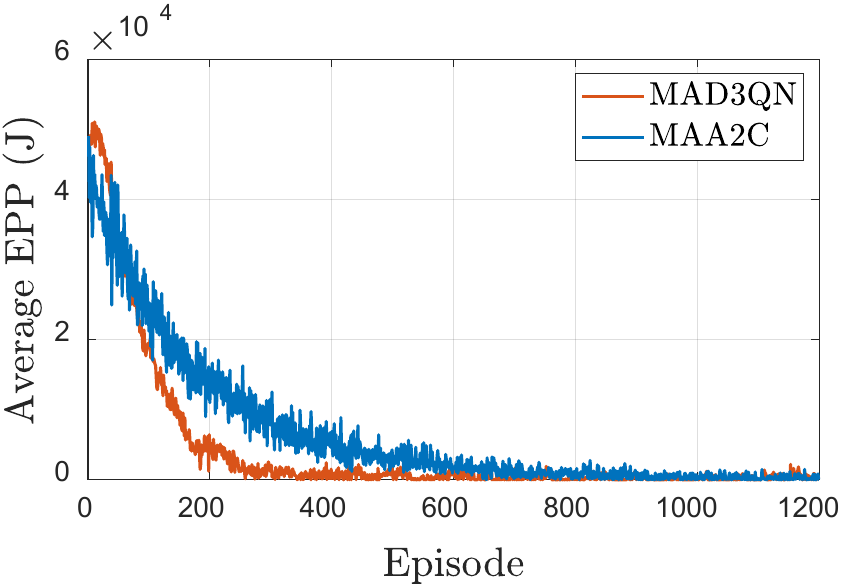}
    \vspace{-3mm}
    \caption{The convergence performance of MAD3QN and MAA2C under $N=5$k EDs.}
    \label{fig_train}
\vspace{-3mm}
\end{figure}

\begin{figure}[t]
    \centering
    \includegraphics[width=3.1in]{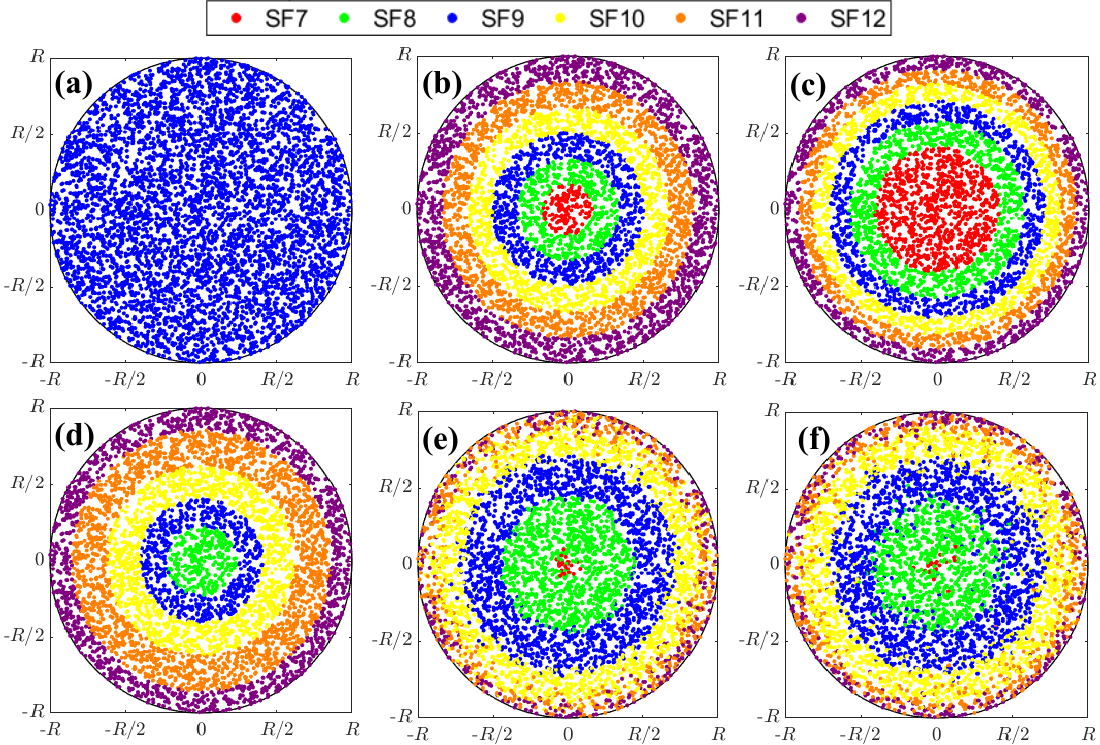}
    \vspace{-3mm}
    \caption{SF distribution of $N=5$k EDs under (a) Same SF, (b) EIB, (c) EAB, (d) PLB, (e) MAD3QN, and (f) MAA2C approaches. The color of point represents the selected SF configuration, e.g, red for SF7.}
    \label{fig_distribution}
    \vspace{-3mm}
\end{figure}

\begin{figure}[t]
    \centering
    \includegraphics[width=2.3in]{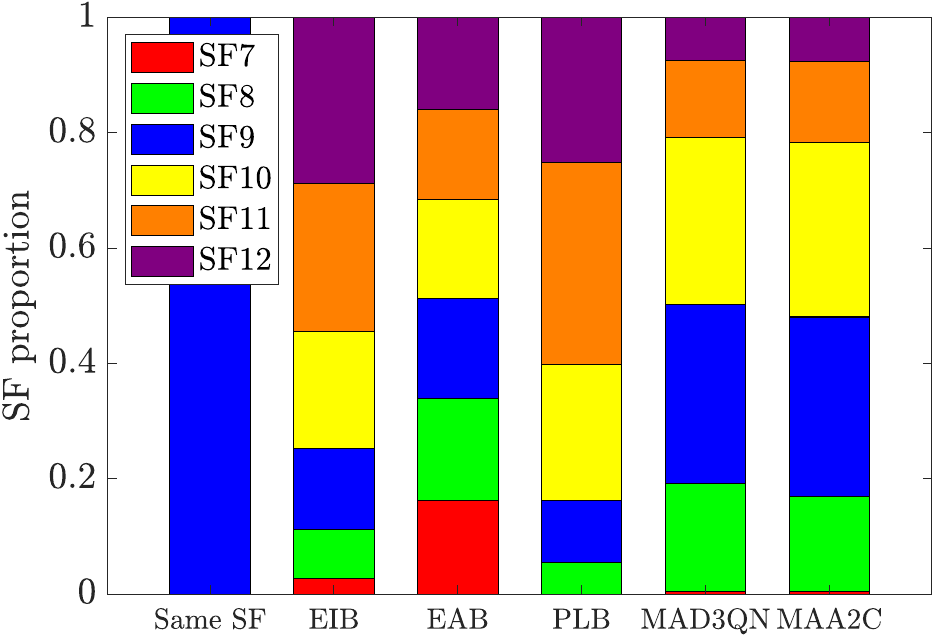}
    \vspace{-3mm}
    \caption{The statistical proportions of each SF for Figs.~\ref{fig_distribution}(a)$\sim$(f), respectively.}
    \label{fig_proportion}
    \vspace{-6mm}
\end{figure}

We first present the average EPP (i.e., $\frac{1}{N}\sum_1^N EPP_n$) under different approaches as a function of the number of EDs $N$ in Fig.~\ref{fig_epp}. One can observe that the average EPP increases with the number of EDs due to the high probability of co-channel co-SF interference in a denser network. Our proposed MAD3QN and MAA2C approaches yield a significant improvement in the average EPP compared to the other four benchmarks, and such a performance gain becomes more pronounced with an increase in $N$. Meanwhile, the average EPP of MAD3QN is slightly lower than that of MAA2C, implying better energy efficiency. Specifically, the average EPP of MAD3QN and MAA2C is 2.46~J and 2.61~J, respectively, at $N=5$k. 

Fig.~\ref{fig_train} depicts the average EPP of MAD3QN and MAA2C under $N=5$k EDs concerning the training episodes, from which we can observe that MAD3QN performs better than MAA2C in terms of convergence rate. For instance, the EPP of MAD3QN converges to a stable value after around $300^{th}$ training episodes, while it takes MAA2C nearly $600^{th}$ training episodes to converge. 

Fig.~\ref{fig_distribution} demonstrates the SFs distribution for $N=5$k EDs under different approaches. In~Fig.~\ref{fig_distribution}(a), the same SF scheme allocates all EDs with SF9, which balances ToA and propagation capability. However, many EDs configured with the same SF can result in a lower $P_{SIR}$. Meanwhile, $P_{SNR}$ of the EDs at the edge degrades due to the limited link budget. Figs.~\ref{fig_distribution}(b) and (d) highlight that EIB and PLB allocate the larger SF (i.e., SF11 and SF12) to the peripheral EDs for a higher $P_{SNR}$. However, the significant number of EDs operating at higher SF levels leads to a considerable increase in EPP. This is primarily due to the elevated transmit power consumption and the higher collision probability caused by the extended ToA. Despite EAB equally assigning SFs among all EDs to mitigate the same-SF interference, as depicted in~Fig~\ref{fig_distribution}(c), the EDs configured with SF7 cannot establish reliable U-DtS connectivity, which results in the worse $P_{SNR}$ and the increased average EPP. Figs.~\ref{fig_distribution}(e) and (f) reveal that our proposed MARL approaches assign the EDs near the inner ring by SF8$\sim$10 while allocating SF11 or SF12 to the EDs at the edge. This allocation strategy aims to accomplish the robust link with the lower transmit power consumption and to reduce the co-SF interference probability, thereby improving the average EPP. Consequently, our proposed approaches exhibit notably superior performance compared to all the four benchmarks. The share of EDs using each SF under the discussed approaches are illustrated in Fig.~\ref{fig_proportion}.

\vspace{-4mm}
\section{Conclusion}
This letter investigates the effectiveness of MARL for optimizing SFs allocation in massive U-DtS scenarios. After developing an analytical model to characterize packet delivery ratio and using it as our reward mechanism, we utilize the MAD3QN and MAA2C approaches to optimize SFs allocation for improving the system's energy efficiency. Through a comparison with the four benchmarks in a realistic farm case, our numerical results reveal that the proposed approaches exhibit the lowest average EPP, where MAD3QN slightly outperforms MAA2C in terms of the average EPP and convergence rate. Note that our proposed MARL approach is universal, and can be generalized for other subterranean LoRaWAN NTN applications. The future work will focus on developing a strategy for broadcasting the derived SF configuration to each ED by considering in more detail the mobility of NTN and the reliable downlink communication~\cite{R2ref1, R2ref2}. 
\vspace{-6mm}

\bibliographystyle{IEEEtran}
\bibliography{ref}

\end{document}